\RequirePackage{fix-cm}
\documentclass[smallextended]{svjour3}       
\smartqed  

\usepackage{graphicx}
\usepackage{multirow}
\usepackage[caption=false]{subfig}
\usepackage{tikz}
\usepackage{xcolor}
\usepackage[multi-part-units=single]{siunitx}
\begin{document}

\title{Needle tip force estimation by deep learning from raw spectral OCT data
}

\titlerunning{Needle tip force estimation by deep learning from raw spectral OCT data}        

\author{M. Gromniak \and N. Gessert \and T.~Saathoff \and A. Schlaefer}


\institute{Martin Gromniak \at
              Institute of Medical Technology, Hamburg University of Technology, Germany\\
              \email{martin.gromniak@tuhh.de}           
}

\date{Accepted for publication in the International Journal of Computer Assisted Radiology and Surgery}

\maketitle
\begin{abstract}
\textit{Purpose} Needle placement is a challenging problem for applications such as biopsy or brachytherapy. Tip force sensing can provide valuable feedback for needle navigation inside the tissue. For this purpose, fiber-optical sensors can be directly integrated into the needle tip. Optical coherence tomography (OCT) can be used to image tissue. Here, we study how to calibrate OCT to sense forces, e.g. during robotic needle placement.

\textit{Methods} We investigate whether using raw spectral OCT data without a typical image reconstruction can improve a deep learning-based calibration between optical signal and forces. For this purpose we consider three different needles with a new, more robust design which are calibrated using convolutional neural networks (CNNs). We compare training the CNNs with the raw OCT signal and the reconstructed depth profiles.

\textit{Results} We find that using raw data as an input for the largest CNN model outperforms the use of reconstructed data with a mean absolute error of 5.81 mN compared to 8.04 mN.

\textit{Conclusions} We find that deep learning with raw spectral OCT data can improve learning for the task of force estimation. Our needle design and calibration approach constitute a very accurate fiber-optical sensor for measuring forces at the needle tip.

\keywords{Optical Coherence Tomography \and Deep Learning \and Force Estimation \and Raw Data}
\end{abstract}


\newpage

\section{Introduction}

Needle placement is a challenging problem for a variety of medical interventions, including brachytherapy or biopsy \cite{taylor2016medical}. The force acting on the needle tip allows for inference about the currently penetrated tissue. This information can be used to navigate the needle and to prevent injuries of delicate structures \cite{okamura2004force}. In order to distinguish tissue based on tip forces, it may be required to measure those with an accuracy of approximately \SI{0.01}{\newton} \cite{Mccreery2008}. Tip forces cannot be measured with external sensors due to friction forces at the needle shaft \cite{kataoka2002measurement}. Therefore, small-scale fiber-optical force estimation methods have been directly integrated into the needle tip. Several sensor concepts are based on Fabry-Pérot interferometry \cite{beekmans2017fiber} and Fiber Bragg Gratings \cite{kumar2016detecting}. Here, we consider a setting where optical coherence tomography is available, e.g., to study tissue deformation \cite{Otte2012} or to realize elastography \cite{Latus2017}. While OCT has been proposed for tip forces estimation before \cite{gessert2018needle,gessert2019spatio}, these approaches rely on the reconstructed gray value data. However, using the reconstructed data has two limitations. First, the signal processing is based on a number of assumptions which may cause some loss of signal information. Second, it does not incorporate the phase part of the complex OCT signal that is particularly sensitive to small axial shifts. Therefore, we explore whether the tip force estimation accuracy can be improved by directly using the raw spectral OCT data. Thus, we perform a calibration between the optical signal and forces applied to the needle tip with convolutional neural networks. We validate our approach with three different needles using a new, improved needle design.

\section{Methods}

\subsection{Needle Design}

\begin{figure}[t]
\begin{tikzpicture}
\node[] at (0,0)
    {\includegraphics[width=\textwidth]{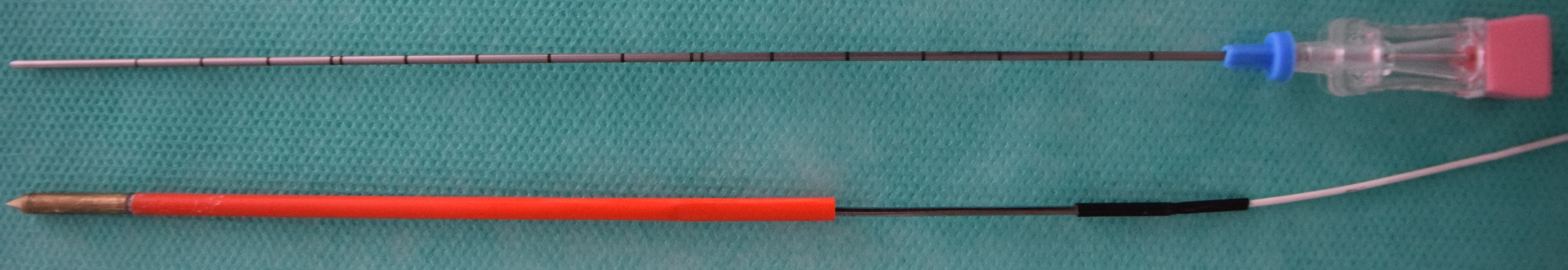} };
\node[] at (-2.3,-2.5)
    {\includegraphics[width=0.6\textwidth]{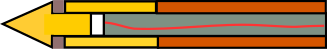}};
\draw[draw=black] (-5.9,-1) rectangle ++(1.5,1);
\draw[draw=black] (-5.9,-3.8) rectangle ++(7.15,2.3);
\draw (-5.9,-1) -- (-5.9,-1.5);
\draw (-4.4,0) -- (1.25,-1.5);

\node at (-4.6,-3) (Epoxy) {};
\node at (-4.6,-3.55) (Epoxy layer) {Epoxy layer};
\draw [->] (Epoxy layer) -- (Epoxy);

\node at (-0.5,-2.45) (Fiber) {};
\node at (-0.5,-3.55) (OCT Fiber) {OCT Fiber};
\draw [->] (OCT Fiber) -- (Fiber);

\node at (-4.9,-1.75) (nodeA) {};
\node at (-4.3,-1.75) (nodeB) {};

\draw [<->] (nodeA) -- (nodeB);

\draw (nodeA) -- (nodeB) node [right] (TextNode) {\SI{0.5}{\milli\metre}};


\end{tikzpicture}

\caption{The design of the needles used in this work. In the upper image one of our needles is depicted below, in comparison to a standard G18 biopsy needle above. In the schema, the brass tip with a piston and the brass sleeve are depicted in yellow. The ferrule guiding the fiber is depicted in gray. The protection tube is depicted in orange.}
\label{fig:needle_design}
\end{figure}
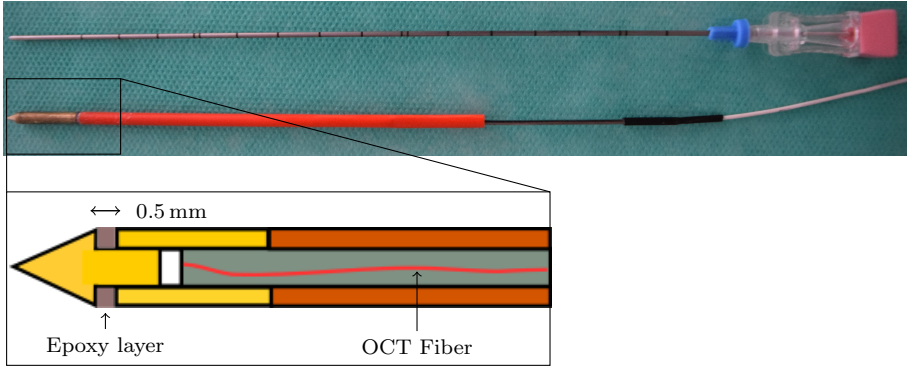

We used an improved needle design for force estimation at the needle tip. A scheme and an image of the needle are shown in Figure \ref{fig:needle_design}. A brass tip with a piston is put on a brass sleeve such that it is able to perform a sliding motion inside of it. An epoxy layer between tip and sleeve acts as a spring. An optical fiber is embedded into a ceramic ferrule. The ferrule is positioned relative to the tip piston such that the light beam travels a distance of approximately $\SI{1}{\milli\metre}$ through air until hitting the surface of the piston. For protection, the ferrule is embedded into a polymer tube which is glued to the brass sleeve. When forces act on the needle tip, the epoxy layer is compressed and the piston moves closer to the exit point of the laser beam which can be detected in the OCT signal. The diameter of the needle is $\SI{2}{\milli\metre}$. 

In \cite{gessert2018needle} the needle tip was constructed as a cone and attached to the needle shaft with a deformable epoxy layer. Thus, radial forces on the needle tip could easily tilt it. The improved piston construction has the advantage that it guides the tip in axial needle direction and prevents tilt. This contributes to a more reproducible signal, an important aspect in the calibration of the needle, and to the overall durability.

\subsection{Calibration Data}

We acquire calibration data for three custom build needles identical in construction. The data acquisition is performed similar to \cite{gessert2019spatio} where a needle is driven against a flat surface with a stepper motor. The force in the axial direction of the needle is measured with a force sensor and recorded together with the associated raw OCT data. Approximately \num{180000} OCT-force pairs are collected for each needle for forces between $\SI{0}{\newton}$ and $\SI{1}{\newton}$. We perform regular OCT data reconstruction, which includes the following steps:

\begin{enumerate}
    \item Dechirping the data by resampling and interpolating to new sampling points, based on manufacturer specifications
    \item Estimation of the DC spectrum using an exponential moving average with a damping coefficient of $d = 0.05$
    \item Subtraction of the DC spectrum
    \item Apodization by using a Hann window for filtering the spectral data
    \item Fourier transform for mapping frequency values to spatial values
    \item Selection of the absolute signal value as the final reconstructed intensity image
\end{enumerate}

\begin{figure}[t]
\centering
\begin{tikzpicture}
\node (raw) at (0,0)
    {\includegraphics[width=.45\textwidth]{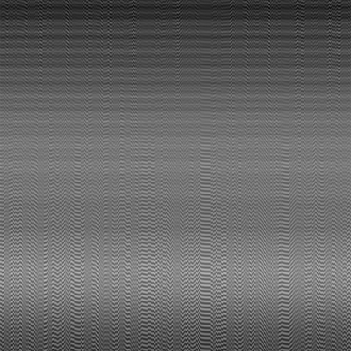}};
\draw[->, line width=0.5mm, color=red] (-2.82,2.8) -- (0,2.8) node [midway, above right] {t};
\draw[->, line width=0.5mm, color=red] (-2.82,2.8) -- (-2.8,0) node [midway, rotate=90, above] {frequency};
\node (raw) at (6,0)
    {\includegraphics[width=.45\textwidth]{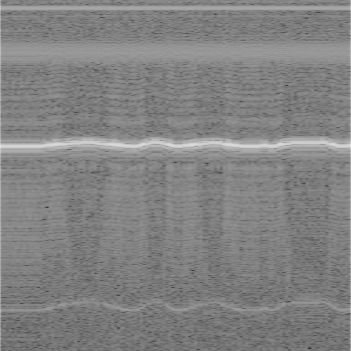}};
\draw[->, line width=0.5mm, color=red] (3.2,2.8) -- (6,2.8) node [midway, above right] {t};
\draw[->, line width=0.5mm, color=red] (3.2,2.8) -- (3.2,0) node [midway, rotate=90, above ] {depth};

\draw[line width=0.5mm, color=red] (7.5,0.45) -- (7,-0.5) node  [ below, fill=white ]  {piston surface};

\end{tikzpicture}
\caption{Example OCT image data over a period of time. Left, the raw OCT signal is shown. Right, the reconstructed A-Scans (depth profiles) are shown. One individual scan is depicted in one column. The scans were collected as part of a time series. Subsequent scans are adjacently depicted (M-Scan). The reconstructed scans allows human interpretation to some extend, e.g., the piston surface is visible. Patterns in the raw OCT data are much more subtle and not interpretable by the human eye.}
\label{fig:raw_reco}
\end{figure}
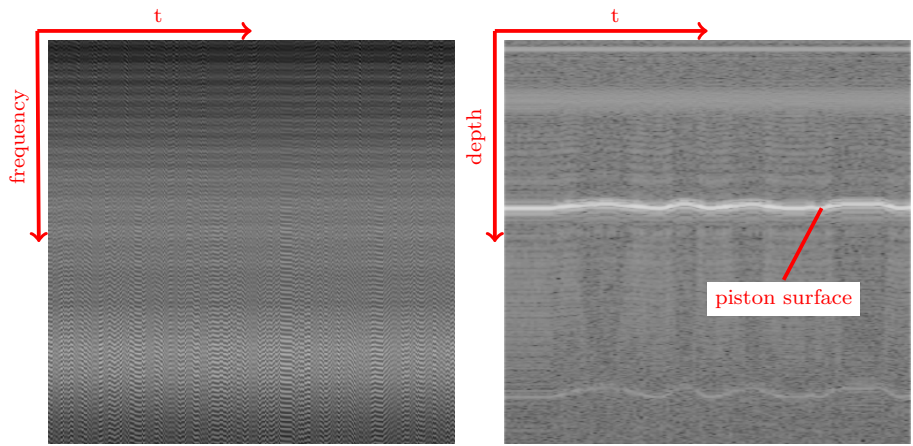

Throughout the reconstruction process, information can be lost due to the DC spectrum estimation strategy (2), the apodization (4) which eliminates high-frequency signal parts and the selection of the absolute signal value (6) as the final image. Figure \ref{fig:raw_reco} shows an excerpt of the collected data, both in raw (left) and reconstructed form (right).

The raw OCT signal has a size of $1024 \times N_t$ where $N_t$ is the number of scans acquired over time. Reconstruction up to step (5) results in a complex signal which is also of size $1024 \times N_t$. Finally, by taking the absolute signal value, the A-Scan image sequence of size $512 \times N_t$ is obtain. This can be interpreted as a sequence of 1D depth images over time.

\subsection{Deep Learning Architectures}

We compare prediction performance for needle tip forces from both raw and reconstructed OCT data with different convolutional neural network (CNN) architectures. The considered architectures are variants of the ResNet \cite{resnet}, which is an extension of CNNs that enables better training through improved gradient flow. The ResNet models were originally built for 2D images with 2D convolutions. Here, the A-Scan data represents 1D images. Therefore, we adapt the architectures by replacing 2D convolutions by 1D convolutions. Furthermore, we replace the network's original output layer for multi-class classification with a fully-connected layer with one output for needle force regression. We consider several ResNet  variants,  resembling  different  network  sizes, the regular ResNet34 and ResNet18 architecture as well as a smaller architecture with 2 residual blocks and 6 convolutional layers, which we name ResNet6. All deep learning models are implemented in PyTorch \cite{paszke2019pytorch}. Learning is performed over 150 epochs with a batch size of $N_B = 128$ and a learning rate of 0.005 using the Adam optimizer. As a loss function we use the mean squared error which is defined as

\begin{equation}
	MSE = \frac{1}{N_B}\sum_{j=1}^{N_B}(y^{j}-\hat{y}^{j})^2
\end{equation}

where $y$ is the ground-truth force and $\hat{y}$ is the predicted force value. We use $\SI{20}{\percent}$ of the data as a hold-out validation set. We performed five training runs with different random seeds and averaged the individual results.

\section{Results}

\begin{figure}
\centering
\includegraphics[scale=0.7]{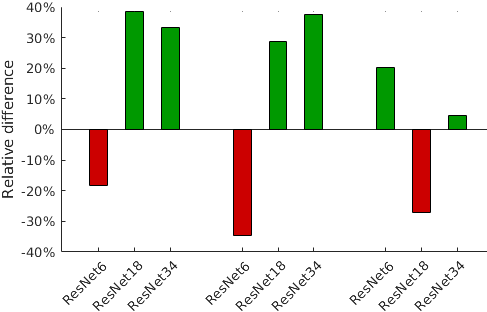}
\subfloat[\label{n1} Needle 1]{\hspace{.22\linewidth}}
\subfloat[\label{n1} Needle 2]{\hspace{.22\linewidth}}
\subfloat[\label{n1} Needle 3]{\hspace{.22\linewidth}}
\caption{Bar plot showing the relative difference of the mean absolute error when learning from raw data.  \label{candleplot}} 
\end{figure}

We report the mean absolute error (MAE) in mN between force predictions and force targets on the validation set. Additionally, we report inference times for the examined neural network architectures. All results are shown in Table~\ref{tab:mae}. Figure \ref{candleplot} shows the relative differences in errors graphically.
For most combinations of needle and model architecture the error for learning on raw OCT data is lower compared with learning from reconstructed OCT scans. Particularly for the ResNet34 architecture, the calibration performance improves for all needles.


\begin{table} [h]
\centering
\begin{tabular}{ |p{1.5cm}||p{0.8cm}|p{0.8cm}||p{0.8cm}| p{0.8cm}|| p{0.8cm}| p{0.8cm} || p{1.6cm}|  }
 \hline
& \multicolumn{2}{|c||}{Needle 1} & \multicolumn{2}{|c||}{Needle 2} & \multicolumn{2}{|c||}{Needle 3} & \multicolumn{1}{|c|}{Inf. times} \\
\hline
& raw & recon & raw & recon & raw & recon &   \\ 
\hline
ResNet 6 &  \hfill $8.54 \newline \pm 0.14$ & \hfill $7.22 \newline \pm 0.14$ & \hfill $23.10 \newline \pm 0.40$ & \hfill $17.15 \newline \pm 0.33$ & \hfill $9.02 \newline \pm 0.08$ & \hfill $11.29 \newline \pm 0.09$ & $1.11 \pm 0.00$ \\ 
 \hline
ResNet 18 & \hfill $4.36 \newline \pm 0.05$ & \hfill $7.09 \newline \pm 0.18 $  & \hfill $8.16 \newline \pm 0.23$ & \hfill $11.42 \newline \pm 0.32$ & \hfill $7.18 \newline \pm 0.66$ & \hfill $5.65 \newline \pm 0.06$ & $3.56 \pm 0.00$\\ 
 \hline
ResNet 34 &  \hfill $4.40 \newline \pm 0.06$ & \hfill $6.61 \newline \pm 0.19$ & \hfill $6.95 \newline \pm 0.24$ & \hfill $11.14 \newline \pm 0.32$ & \hfill $6.08 \newline \pm 0.06$ & \hfill $6.37 \newline \pm 0.05$  & $6.43 \pm 0.00$ \\ 
\hline

\end{tabular}
\centering
\caption{Mean absolute error results in mN and inference times in ms. \label{tab:mae}}
\end{table}

\section{Discussion and Conclusion}

In this paper we address the calibration problem of OCT-based needle tip force estimation using new and improved sensor concept with a piston and a guiding sleeve. In contrast to a previous OCT-based concept \cite{gessert2018needle}, the needle is not as sensitive to lateral forces by design while achieving similar calibration results with intensity data.

Furthermore, we study an approach for improving deep learning-based calibration performance even further. During OCT acquisition a spectral signal is obtained which is typically reconstructed to an intensity depth image.  We illustrate that learning is possible in an end-to-end fashion, i.e., the process of image reconstruction can be avoided. Moreover, the results improve in six out of nine setups, indicating that there may be information lost in the original processing that can be used when training the network on the raw signal. The proposed approach allows for precise estimation of forces at the needle tip, which is particularly interesting for force based robotic needle placement. With typical robot control cycles of \SI{1}{\milli \second}, a trade-off between accuracy and inference time must be balanced.

We find that learning forces from raw OCT data instead of reconstructed images work well, in particular, for larger deep learning models. For future work, our approach could be studied in more detail with additional deep learning methods and in different applications scenarios. Also, our approach could be extended to other applications where an imaging modality is used as a sensor signal, for example in the context of motion tracking with OCT or ultrasound.

\section*{Compliance with Ethical Standards}

\small \textbf{Funding:} This work was partially supported by the TUHH $i^3$ initiative and DFG grants SCHL 1844/2-1 and SCHL 1844/2-2.

\small \textbf{Conflict of Interest:} 
The authors M. Gromniak, N. Gessert, T. Saathoff and A. Schlaefer declare that they have no conflict of interest.

\small \textbf{Ethical Approval:} 
This article does not contain any studies with human participants or animals.

\small \textbf{Informed Consent:} 
Not applicable.


%
%

\bibliographystyle{spmpsci}      
\bibliography{references.bib}   

\begin{thebibliography}{10}
\providecommand{\url}[1]{{#1}}
\providecommand{\urlprefix}{URL }
\expandafter\ifx\csname urlstyle\endcsname\relax
  \providecommand{\doi}[1]{DOI~\discretionary{}{}{}#1}\else
  \providecommand{\doi}{DOI~\discretionary{}{}{}\begingroup
  \urlstyle{rm}\Url}\fi

\bibitem{taylor2016medical}
Taylor, R.H., Menciassi, A., Fichtinger, G., Fiorini, P., Dario, P. (2016)
  Medical robotics and computer-integrated surgery.
\newblock In: Springer handbook of robotics, pp. 1657--1684. Springer

\bibitem{okamura2004force}
Okamura, A.M., Simone, C., O'leary, M.D. (2004) Force modeling for needle
  insertion into soft tissue.
\newblock IEEE transactions on biomedical engineering \textbf{51}(10),
  1707--1716

\bibitem{Mccreery2008}
Mccreery, G., Trejos, A., Naish, M., Patel, R., Malthaner, R. (2008)
  Feasibility of locating tumours in lung via kinaesthetic feedback.
\newblock The international journal of medical robotics and computer assisted
  surgery : MRCAS \textbf{4}, 58--68.
\newblock \doi{10.1002/rcs.169}

\bibitem{kataoka2002measurement}
Kataoka, H., Washio, T., Chinzei, K., Mizuhara, K., Simone, C., Okamura, A.M.
  (2002) Measurement of the tip and friction force acting on a needle during
  penetration.
\newblock In: MICCAI, pp. 216--223. Springer

\bibitem{beekmans2017fiber}
Beekmans, S., Lembrechts, T., van~den Dobbelsteen, J., van Gerwen, D. (2017)
  Fiber-optic fabry-p{\'e}rot interferometers for axial force sensing on the
  tip of a needle.
\newblock Sensors \textbf{17}(1), 38

\bibitem{kumar2016detecting}
Kumar, S., Shrikanth, V., Amrutur, B., Asokan, S., Bobji, M.S. (2016) Detecting
  stages of needle penetration into tissues through force estimation at needle
  tip using fiber bragg grating sensors.
\newblock Journal of biomedical optics \textbf{21}(12), 127,009

\bibitem{Otte2012}
Otte, C., H{\"u}ttmann, G., Schlaefer, A. (2012) Feasibiliy of optical
  detection of soft tissue deformation during needle insertion.
\newblock In: SPIE 8316, Medical Imaging 2012: Image-Guided Procedures, Robotic
  Interventions, and Modeling, vol. 8316, pp. 282 -- 292

\bibitem{Latus2017}
Latus, S., Otte, C., Schl{\"u}ter, M., Rehra, J., Bizon, K.,
  Schulz-Hildebrandt, H., Saathoff, T., H{\"u}ttmann, G., Schlaefer, A. (2017)
  An approach for needle based optical coherence elastography measurements.
\newblock MICCAI 201: Medical Image Computing and Computer-Assisted
  Intervention pp. 655--663

\bibitem{gessert2018needle}
Gessert, N., Priegnitz, T., Saathoff, T., Antoni, S.T., Meyer, D., Hamann,
  M.F., J{\"u}nemann, K.P., Otte, C., Schlaefer, A. (2018) Needle tip force
  estimation using an oct fiber and a fused convgru-cnn architecture.
\newblock In: MICCAI, pp. 222--229. Springer

\bibitem{gessert2019spatio}
Gessert, N., Priegnitz, T., Saathoff, T., Antoni, S.T., Meyer, D., Hamann,
  M.F., J{\"u}nemann, K.P., Otte, C., Schlaefer, A. (2019) Spatio-temporal deep
  learning models for tip force estimation during needle insertion.
\newblock Int J CARS \textbf{14}, 1485–1493

\bibitem{resnet}
He, K., Zhang, X., Ren, S., Sun, J. (2016) Deep residual learning for image
  recognition.
\newblock pp. 770--778.
\newblock \doi{10.1109/CVPR.2016.90}

\bibitem{paszke2019pytorch}
Paszke, A., Gross, S., Massa, F., Lerer, A., Bradbury, J., Chanan, G., Killeen,
  T., Lin, Z., Gimelshein, N., Antiga, L., Desmaison, A., Kopf, A., Yang, E.,
  DeVito, Z., Raison, M., Tejani, A., Chilamkurthy, S., Steiner, B., Fang, L.,
  Bai, J., Chintala, S. (2019) Pytorch: An imperative style, high-performance
  deep learning library.
\newblock In: Advances in Neural Information Processing Systems 32, pp.
  8024--8035. Curran Associates, Inc.

\end{thebibliography}

\end{document}